\documentclass[final,1p,times,twocolumn]{elsarticle}

\usepackage{graphicx}

\usepackage[style=list,number=none]{glossary}
\noist \makeglossary

\journal{Trends in Ecology and Evolution}

\begin{document}

\begin{frontmatter}



\title{The contribution of statistical physics to evolutionary biology}

 \author[ista]{Harold P. de Vladar\corref{cor1}}
  \ead{hpvladar@ist.ac.at}
  \cortext[cor1]{Corresponding author}
 \author[ista,ued]{Nicholas H. Barton}
  \ead{Nick.Barton@ist.ac.at}
  \address[ista]{I.S.T. Austria. Klosterneuburg A-3400, Austria}
\address[ued]{Institute of Evolutionary Biology, University of Edinburgh. Edinburgh, United Kingdom}

\begin{abstract}
Evolutionary biology shares many concepts with statistical physics: both deal with populations, whether of molecules or organisms, and both seek to simplify evolution in very many dimensions. Often, methodologies have undergone parallel and independent development, as with stochastic methods in population genetics. We discuss aspects of population genetics that have embraced methods from physics: amongst others, non-equilibrium statistical mechanics, travelling waves, and Monte-Carlo methods have been used to study polygenic evolution, rates of adaptation, and range expansions. These applications indicate that evolutionary biology can further benefit from interactions with other areas of statistical physics, for example, by following the distribution of paths taken by a population through time.
\end{abstract}


\end{frontmatter}


\section*{Parallel foundations of evolution and statistical physics}
\label{sect:intro}

In the late 19th century, Boltzmann established the theoretical foundations of {\bf statistical mechanics}, in which the behaviour of ensembles of particles explains large-scale phenomena \cite{Boltzmann1896}. For example, the position and velocity of the particles in a gas can fluctuate between very many states (termed micro-states), but averages over all the configurations that give the same observable macroscopic states (temperature and pressure, say) \cite{Callen1985}. A similar averaging over equivalent micro-states is made in both population and quantitative genetics: we average over individual gene combinations to describe a population by its allele frequencies, and we can further average over all the allele frequencies that are consistent with a given mean and variance of a quantitative trait. In this sense, physicists and evolutionary biologists both model populations (a gas or a gene pool) rather than precise types (individual particles or  genotypes). This ``statistical'' description in terms of a few variables, the macro-states, summarizes the many possible configurations of the micro-states (degrees of freedom), which cannot be accurately measured or described. Furthermore, the macro-states are then sufficient to predict other properties without reference to the micro-states. For example, thermodynamics describes macroscopic properties without referring to individual particles; similarly, quantitative genetics does not refer to allele frequencies to predict the trait mean in the next generation.

Hence, evolutionary biology and statistical physics often use similar theoretical methodologies, although studying very different phenomena. We argue that there are close analogies between evolutionary genetics and statistical physics. Physical techniques had an early influence on molecular biology (\ref{Box:History}). But more specifically, non-equilibrium methods are based on the same theory of stochastic processes that is used in population genetics. Thus, some physical theories promise further developments that can deepen our understanding of evolution in two ways: either by applying common mathematical techniques (e.g. diffusion equations, see \ref{Box:DiffEq}), or by developing precise analogies that incorporate new concepts (e.g. ensemble averaging, information, or {\bf entropy}).  These techniques are being applied in different aspects of evolutionary biology. This article focuses mainly in those that we consider most promising for population genetics. We aim to introduce to the reader to these methods by reviewing representative examples in the literature.

\section*{The cost of selection, entropy and information}
\label{sect:SelectionEntropyInformation}
It is extraordinary that the selection of random mutations has created complex organisms that appear exquisitely designed to fit their environment. Selection can be seen as taking information from the environment, and coding it into the DNA sequence \cite{MaynardSmith:2000p4611}: thus, the gene pool contains information about those specific sequences that confer high fitness. This idea can be quantified using the concepts of entropy and information \cite{Shannon:1948p4599}. Entropy is a measure of the number of different states in which a population is likely to be found: thus, selection of one specific genotype, or genotype frequency, corresponds to minimal entropy. A asking how the genotype of an individual, or a population, depends on the selection that they have experienced, can be quantified by an entropy that measures how strongly selection has clustered the population around a specific genotype. 
Haldane \cite{Haldane:1957p213} showed that the number of {\bf selective deaths} needed to fix an allele is independent of the selection pressure, and Kimura \cite{Kimura:1961p5973} pointed out that Haldane's ``cost of natural selection'' is exactly the information gained by fixing a specific allele. This relation applies very generally to asexual populations \cite{Worden:1995fk} but fails with recombination (see below). The theory of {\bf quasispecies} (a model of mutation-selection balance), emphasizes that the reproductive rate (selection) limits the amount of information that can be maintained in the face of random mutations \cite{Eigen:1977p4220}. However, this constraint can be relaxed with recombination and epistasis \cite{Kimura:1966p7222}. Analogies with statistical physics help us to understand how selection accumulates information. We first consider infinite populations -- evolving deterministically -- and then the more general case where random drift in finite populations drives evolution. 

\subsection*{Deterministic evolution and the role of recombination}
The information content of a single large population that is evolving deterministically is measured by the entropy, defined as \(S=\sum_x p_x \log(p_x)\)  where $p_x$ is the frequency of alleles or genotype $x$ \cite{Haldane:1957p213,Eigen:1977p4220}. This entropy reflects the information accumulated and maintained by evolution, and is closely related to Shannon's information [3-4]. Remarkably, the {\bf replicator dynamics} can be obtained by maximizing a different measure, {\bf Fisher information}: \(F=\sum_x p_x \left(\frac{d}{dt}\log(p_x) \right)^2\) , which measures divergence between two distributions \cite{Frieden:2001p4597,Frank:2009p4318}. Fisher's information is the ``acceleration'' of the entropy, i.e. \(d^2S/dt^2=F \), where we interpret \(\frac{d}{dt}\log(p_x)\)  as the information contained about selection when we observe how the frequency of $x$ has changed. For example, for a beneficial allele under selection, this would be proportional to the selective value $s$.  Normally, we predict the change of the frequency distribution when we know $s$. Fisher's information takes the parent and the offspring distributions as given, and measures the effect of selection from the difference between these two \cite{Frank:2009p4318}.

\paragraph{\bf Sexual vs. asexual reproduction} It has long been understood that, when combined with {\bf truncation selection}, sexual reproduction is much more efficient than asexual in fixing beneficial genotypes \cite[Ch. 2]{Crow:1964uq,Ewens:1979}. In the former case, the maximum information increases as $n^{1/2}$ ($n$ being the number of loci), while in the latter, it increases by only  
one unit per generation \cite{MaynardSmith:2000p4611,MacKay2003}. The maximum number of loci that can be maintained despite the randomizing effect of mutation is $~1/\mu$ for asexuals \cite{MacKay2003}, whilst for sexuals (with free recombination) it can be as high as $~1/\mu^2$, where $\mu$ is the mutation rate at each locus \cite{MacKay2003,Peck:2010p6334}. (More loci would produce more mutants, and hence decrease the amount of information). According to Haldane's principle \cite{Haldane:1937kx} every deleterious mutation must be eliminated by a failure to reproduce (a ``selective death''). Therefore, the mutation load is independent of selection strength, and is half as great if selection eliminates two copies in a recessive homozygote at the same time. In haploids, redundancy leads to a similar gain in efficiency \cite{Peck:2010p6334,Watkins:2002vn}.

\section*{Stochastic evolution: the diffusion of allele frequencies}
\label{sect:StochasticEvolution}
The diffusion approximation shows how the distribution of allele frequencies at many loci changes through time in finite populations. In this case, selection, mutation and migration are modelled as deterministic factors, and genetic drift introduces random fluctuations to populations within an ensemble (\ref{Box:DiffEq}). (Other treatments are possible, where mutations are regarded as carrying random changes to individuals within a single population.)
In other fields, constant diffusion coefficients have been widely used, leading to simple Gaussian solutions (\ref{Box:DiffEq}). However, Gaussian solutions are not appropriate for population genetics, because allele frequencies range between zero and one, and sometimes cluster near fixation, in a bimodal distribution. After its introduction by Fisher \cite{Fisher:1922lh}, Kolmogorov applied the more general diffusion method to the neutral island model \cite{Kolmogorov:1935zr}, which Wright \cite{Wright:1931} had already solved by different means. Kimura relied on the diffusion approximation to model the evolution of finite populations \cite{Kimura:1955ly}, and for his neutral theory of molecular evolution \cite{Kimura:1985}. 

The diffusion approximation, central to both population genetics and statistical physics,  provides a way to model many factors in a mathematically tractable way. Crucially, it approximates a wide variety of more detailed models. Mathematically, it is equivalent to the coalescent process that describes the evolution of samples from a population, and to path ensemble methods that describe the distribution of population histories (see below). In physics, diffusion equations describe non-equilibrium processes and are hard to relate to quantities like temperature, entropy, or free energy, which are well-defined only in thermodynamic equilibrium through the {\bf Boltzmann distribution.}

Wright showed that selection, mutation and drift give an explicit distribution, proportional to \(\bar{W}^{2N}\) , where  \(\bar{W}\)  is the mean fitness of a population of size $N$ \cite{Wright:1931}. This is closely analogous to the Boltzmann distribution \( \left( \sim e^{-E/kT} \right) \) \cite{Boltzmann1896,Callen1985}, with  $\log(\bar{W})$ corresponding to (negative) energy, $-E$, and $1/2N$ to the temperature, $kT$ (\ref{Box:DiffEq}). This result was the basis for Wright's metaphor of an adaptive landscape: a surface of mean fitness laid over the multidimensional space of allele frequencies \cite{Wright:1988p1226} (\ref{Box:DiffEq} and Fig. \ref{Fig:DESols}). 

\paragraph{\bf Jumps between adaptive peaks} When the {\bf stationary distribution} is clustered around alternative peaks in the adaptive landscape, the rate at which random drift causes shifts between these states is approximated by a general formula that is proportional to the probability of being at the saddle point (adaptive valley) that separates them, and to the leading eigenvalue that describes the instability at that point \cite{Barton:1987p4186,Wright:1941ve}. Wright \cite{Wright:1941ve} worked out transition rates for chromosome rearrangements, ideas rigorously formulated later using diffusions \cite{Lande:1985qf}. Rouhani and Barton\cite{Barton:1987p4186,Rouhani:1987bh} found the rate of peak shifts in a spatially structured population, borrowing from an identical analysis of transitions between alternative vacuum states.

\paragraph{\bf Traveling waves} The distribution of a quantitative trait, or of fitness itself, can be seen as a {\bf traveling wave} that travels at a steady rate as the population adapts, either in actual or in phenotypic space. Most analyses have been of asexuals, which increase their fitness by accumulation of favourable mutations, or decline under Muller's ratchet \cite{Wilke:2004dq,Rouzine:2007p3220,Burger:1999cr}. Beneficial mutations increase in frequency independently at the wave front, where frequencies are low and subject to drift, but the rest of the wave follows deterministically \cite{Hallatschek:2011p7231}. The wave thus moves at a velocity proportional to the mutation rate, and which depends logarithmically on the population size because of strong random drift at the leading edge \cite{Wilke:2004dq}. This approach has been extended to low rates of recombination \cite{Rouzine:2010nx,Neher:2010oq}. With sexual reproduction, random drift has much less effect, and the population adapts much more quickly \cite{Burger:1999cr,Peck:1999kl}. However, when there is a very high rate of substitution and recombination, {\bf Hill-Robertson interference} limits adaptation rate \cite{Barton:2009tg}.

\paragraph{\bf Spatial evolution and range expansions} Fisher introduced a simple non-linear diffusion equation describing the spread of a beneficial mutation through space \cite{Fisher:1937p5980}. Though motivated by an evolutionary problem, this model raised interest among physicists and mathematicians (establishing a sub-discipline studying the \emph{Fisher-KPP model} --for Kolmogorov-Piskounov-Petrovski, co-discoverers of the model \cite{Kolmogorov:1991fk}). Travelling waves explain the decreased genetic diversity that arises from hitchhiking at the leading edge \cite{Hallatschek:2008hc,Ralph:2010ij}. This approach also provides a practical way to measure selection coefficients \cite{Bauer:1989jl}, and perhaps, a means to distinguish fixation due to selective sweeps from simple drift \cite{Korolev:2010gb,Hallatschek:2010bs}.

\section*{Statistical mechanics and the quantitative genetics of finite populations}
\label{sect:StatMech}
Although the diffusion equation provides an exact description of evolution, the joint distribution at many loci is hard to grasp. Statistical mechanics simplifies the problem by following just a few variables that summarize all the allele frequencies (or in physics, the particles' states). These map the fitness landscapes for allele frequency onto a simpler one for quantitative traits \cite{Lande:1976fv}, which are analogous to macroscopic quantities in statistical physics (\ref{Box:DiffEq}, Fig. \ref{Fig:SimpsonLandscape})\cite{Rattray:2001p19}.

\paragraph{\bf Maximization of entropy} This reduction in dimensionality requires a way to account for the degrees of freedom lost in averaging over the underlying genetic states. This can be achieved by applying the principle of entropy maximization: we assume that the unknown micro-states follow a distribution that maximizes their entropy, $S$, given the values of macroscopic quantities \cite{PrugelBennett:1997p202}. Entropy can be defined in several ways. The definition appropriate here is analogous to the above, but extends to the case when $\underline{p}$ (the vector of allele frequencies at each locus) is the random variable:  \(S=-\int \psi \log[\psi/\varphi]d\underline{p} \). This defines the dispersion of the distribution of allele frequencies, $\psi$, relative to a base distribution, $\varphi$: it is maximized when the distribution is selectively neutral (\(\psi=\varphi\)) and decreases as the distribution becomes more tightly clustered around states that are a priori improbable \cite{Barton:2008p226,Barton:2009p952,Iwasa:1988p12}.
If $S$is maximized whilst constraining the expectations of some macroscopic variables,\(\langle A_i \rangle= \int A_i \psi d\underline{p}\), we obtain a distribution of allele frequencies \(\psi = Z^{-1} \varphi \exp[2N \sum_i \alpha_i A_i]\) \cite{Barton:2008p226}, where $Z$ normalizes the distribution and $N$ is the population size. Remarkably, this distribution corresponds exactly to the stationary solution of the diffusion equation (\ref{Box:DiffEq}), when the $A$'s are chosen according to the particular mode of selection (quantitative traits, genetic variance, etc.) and heterozygosity, and are conjugated with the $\alpha$'s , which are the selection coefficients, mutation rates, etc. \cite{Barton:2008p226,Barton:2009p952,Iwasa:1988p12}. This analogy between statistical mechanics and evolution of a finite population has yielded several results, of which we will mention a few.

The dynamics of polygenic evolution can be approximated by a quasi-equilibrium assumption, that is, that the transient distribution of allele frequencies behaves as if the entropy is maximized at all times, given the current values of macroscopic variables. In this way, the change through time of quantitative characters -- including their genetic variance -- can be computed for populations affected by mutation, selection and drift, for an arbitrary number of loci \cite{Barton:2008p226,deVladar:2011p5333}. In physics, macroscopic systems often change far more slowly than the microscopic fluctuations, justifying this approximation. In biology, we do not have such a stark separation. But nevertheless, the approximation is remarkably accurate even when the environment changes abruptly \cite{Barton:2008p226,Barton:2009p952,deVladar:2011p5333}; traveling waves may provide an explanation \cite{Hallatschek:2011p7231}

\paragraph{\bf Adaptive landscapes and detailed balance} Wright's formula for the stationary distribution \cite{Wright:1931} requires {\bf detailed balance} \cite{Sella:2005p1584}. Population geneticists have shown that detailed balance is generally violated when there are more than two alleles at a locus \cite{Wright:1931}, when recombination or migration are comparable with the strength of selection, or under frequency-dependent selection \cite{Taylor:2006p7233}. Without detailed balance, the dynamics cannot be represented by an adaptive landscape, and can mathematically intractable (though see \cite{Ao:2008p1853}). Phylogenetic analysis reveals deviations from detailed balance -- for example, when genomic GC content changes over time \cite{Galtier:2001p7165}. So, we need methods for analyzing populations that are in a stationary state that violates detailed balance, or that are not at a statistical equilibrium at all. 

\paragraph{\bf Path ensembles} An alternative method that holds without detailed balance is the{\bf path ensemble} \cite{Barton:1987p4186,Mustonen:2010p5306}. Instead of describing the distribution of allele frequencies at any single time, we follow the distribution of paths of allele frequencies between two states at different time-points (Fig. \ref{Fig:FitnessFlux}). The probability of any path can be written down in a simple form, and the chance of a transition from one state to another obtained (in principle) by integrating over paths (\ref{Box:DiffEq}). The trajectories are weighted with respect to an optimal one, through three terms: Fisher's information, the variance in fitness, and the {\bf fitness flux}, $\phi$  (\ref{Box:PathEnsemble}) \cite{Mustonen:2010p5306}. The latter measures the net amount of adaptation given a population's history. It is defined as \(\phi=s \frac{dp}{dt}\) , where $s$ is the selective coefficient of the beneficial allele; $\phi$ is the increase in mean fitness that is expected from changes in allele frequency --but without allowing for changes in selection. The fitness flux is distinct from the change in mean fitness, which in general is not well-defined when selection changes through time. Fitness flux includes changes in allele frequencies due to all evolutionary processes, and to the extent that these interfere with selection, can be negative.
In considering the history of a population, the path ensemble methods give an understanding of the adaptation and evolution of complex traits that accounts for historical contingencies, an advantage over models that only consider a population's state at a given time. 

\section*{Evolutionary biology and Monte-Carlo methods}
\label{sect:EvolBiolMC}
Monte Carlo methods are now widely used in statistical inference. When many variables are involved it is not feasible to explore the whole space of possible states (e.g. all possible phylogenetic trees amongst multiple species). A group working on nuclear weapon development at Los Alamos introduced a simple but widely used algorithm \cite{Metropolis:1953p4288}. One simply makes a random change to the microscopic variables, accepting it if it increases some measure, $L$ (for example, mean fitness). Changes that decrease $L$ to $L^*$, are accepted with probability $L^*/L$. This ensures that the microscopic variables will follow a distribution proportional to the stationary distribution of the diffusion equation, which would in turn, be determined solely by the random changes, multiplied by $L$ (Fig. \ref{Fig:DESols}). This \emph{Metropolis algorithm} has been developed in a statistical context \cite{Hastings:1970p4287}, and applied to generate likelihood surfaces for statistical inference \cite{Szymura:1986dz,Geyer:1992fu,Beaumont:2010uq}. Intriguingly, this algorithm uses a simple form of selection to generate a distribution equal to the product of a neutral base distribution, and the measure $L$ -- just as selection and random drift lead to Wright's distribution under the diffusion approximation (see \ref{Box:DiffEq}). Both rely on detailed balance, but a path ensemble approach allows extension to more general cases\cite{Barton:1987p4186}.

\section*{Obstacles to overcome}
\label{section:obstacles}
\paragraph{\bf Toy models and method-oriented analyses} Over the last decade, physicists have shown strong interest in evolution. For example, in the last five years, over 2000 publications on evolution appeared in physics journals (chiefly \emph{Physical Review} journals, \emph{Physica A}, and \emph{PNAS}). Unfortunately, most of these works pay little attention to the fundamental biology, because the motivation is often the specific methods rather than the biological questions. Consequently, many of these contributions remain unconnected to the rest of the evolutionary theory; for the most part, there is very little communication between the disciplines. Two examples follow.
In the \emph{Bak-Sneppen model} \cite{Bak:1993kl}, populations evolve by removing the least fit individual together with two unrelated neighbours, and replacing them by three new individuals with random fitness. A ``critical value'' is reached, but with repeated periods where the fitness distribution spreads, and then re-organizes to the critical value. The Bak-Sneppen model attempted to explain the distributions of extinction episodes \cite{SNEPPEN:1995p5544}, and patterns of experimental evolution \cite{Elena:2005qa}, but had little impact in biology because it lacks any mechanistic basis. Notably, only 13 of ~700 citations of the Bak-Sneppen model  \cite{Bak:1993kl} were by non-physicists. 
Second, in the \emph{Penna bit-string model of ageing} \cite{Penna:1995mi}, the position in the genome of an allele dictates the age at which its detrimental effect is expressed. A threshold for the total number of such deleterious mutations is set arbitrarily, and the population evolves under mutation and competition. Senescence arises because selection is less effective in late life --a phenomenon already well-understood from Hamilton's general analysis \cite{Hamilton:1966zt}. Here, out of roughly 230 to ref. \cite{Penna:1995mi}, only 5 did not include physicists. These two approaches, and others alike, are not taken seriously since they rest on "toy models" that are not connected with biological reality.

\paragraph{\bf Two problems that restrict communication between disciplines} First, the language and nomenclature employed by physicists are often not consistent with basic concepts in genetics: they employ terms such as energy, spin glass, magnetization, Ising chain, etc. where they should use mean fitness, polygenes, directional selection, or polygenic trait \cite{Baake:1997ys,Baake:2001p5991,Barton:2009p952,Hermisson:2002pi}. Standard population genetics notation is largely ignored, making even the most basic equations appear unfamiliar. To take a central example, the diffusion equation includes deterministic and stochastic ``forces''. In evolution, the stochastic part models genetic drift. However, the term ``drift'' is used in physics to refer to the deterministic part! Different nomenclatures make it difficult for physicists to address important biological questions, and for biologists to understand the questions posed by physicists. This is amplified when new ideas are introduced. For example, in an explanation of the advantages of sex, the idea of mixability was introduced \cite{Livnat:2010p5546}: i.e. sex favours alleles that are fit across different genetic backgrounds. A recently proposed measure of ÔÔmixability'' \cite{Livnat:2010p5546} is identical to Fisher's analysis of variance, which was devised precisely as a measure of epistasis \cite{Fisher1918}. Take another example: a statistical mechanics approach was used to find the distributions of contributions made by individual ancestors to future generations. This defined the statistical ``weight'' of each individual's contribution in a lineage \cite{Derrida:2000ff}, which, in biological terms, is just the reproductive value of an individual -- again, a concept introduced by Fisher \cite{Fisher30}.

Second, known results are often rediscovered due to the lack of a common language. For example, the original result that {\bf free fitness} increases in evolution was illustrated with several examples from population and quantitative genetics, and was interpreted in terms of selection and drift \cite{Iwasa:1988p12}. Yet, the same principle was twice rediscovered by physicists decades later but with more restricted scope \cite{Sella:2005p1584} Another example is the \emph{NK model}, where the fitness landscape can be``tuned", altering the degree of epistasis for fitness, was used to show that recombination is an evolvable trait \cite{Kauffman1993}. Yet, the theoretical analysis of the evolution of sex and recombination has been a thriving field since the 1970's \cite{MaynardSmith1978}. No doubt population geneticists have re-derived results well known in physics (e.g. Wright's calculation of rates of shift between adaptive peaks), but these are not usually published as new physics, and are typically studied for their biological implications.
Nevertheless, physicists have also had a serious commitment to subjects meaningful to evolution. Significant works include those discussed in this article, clonal interference in asexuals \cite{Rouzine:2007p3220,Rouzine:2010nx,Neher:2010oq,Hallatschek:2008hc}, an application of percolation theory to speciation \cite{Gavrilets:1997p155}, extending Haldane's principle to a multilocus trait with partial dominance, epistasis and sexual reproduction \cite{Baake:1997ys,Baake:2001p5991}, and ecological explanations of replicator dynamics \cite{Demetrius:1983kx,Demetrius:2007p7235}. All these are aimed directly at a biological audience, published in appropriate journals. Generally, physicists often have a sharp intuition about their models, which greatly helps in finding solutions. 

Statistical physics is based on universal physical laws. In contrast, biological concepts are relative, plastic, or even arbitrary (e.g. mean fitness, traits). Hence the analogies with statistical-mechanical models are limited, depending on the nature of epistasis, physical linkage of the genes, unpredictable fluctuating selection, etc. Moreover, there are different ways in which precise analogies can be drawn, limiting their scope: some factors act deterministically (e.g. selection) and other stochastically (mutations or drift).

\section*{Conclusions}
\label{setc:conclusions}
Many of the fundamental processes of both population genetics and statistical physics are described by diffusion. In evolution, it provides a common framework for features such as the change in allele frequencies \cite{Wright:1931,Wright:1937p5322}, genealogies  \cite{Barton:2004p4637}, and spatial dispersal \cite{Fisher:1937p5980}. All these, and others, can benefit from methods of non-equilibrium statistical mechanics, which is a major and active field in physics.

The concept of a path ensemble is especially useful, shifting the paradigm from tracking frequencies at each point in time, to considering selection over the whole history of the alleles \cite{Barton:2004p4637}. This can be applied to both, deterministic \cite{Leibler:2010p6108} and stochastic evolution \cite{Barton:1987p4186}. In turn, long-standing questions about the efficiency of natural selection in building complex phenotypes \cite{MaynardSmith:2000p4611,Haldane:1957p213,Kimura:1961p5973}, and evolution under fluctuating selection, can be re-addressed.

Of course, we can ask whether the mathematical paraphernalia that we advocate is of any practical use. Although we should not take mathematical models too literally, they are useful both for generating hypotheses about evolution, and for making sense of ecological and genetic data. Most notably, the neutral theory provides the conceptual framework for analyses of sequence data \cite{Kimura:1985}, and quantitative genetics predicts the effects of selection on complex traits \cite{LynchWalsh:1998}. Ideas from statistical mechanics may help by providing new ways to describe the evolution of complex traits, and by suggesting constraints on the efficacy of selection. A clearer understanding of concepts such as fitness flux and entropy suggest new ways to think about the evolution of quantitative traits. To understand adaptation, we need to contemplate not only the current state of populations, but also their history. This is of course an old idea, but the rationale that we review, suggest new ways to understand the process of adaptation in a historical and quantitative way.  

\paragraph{\bf Acknowledgments} We would like to thank J.P. Bollback, R. Cipriani, J. Hermisson, J. Polechova, and D. Weissman for their comments and observations. This research was funded by the ERC-2009-AdG Grant for project 250152 SELECTIONINFORMATION.

\newpage

\glossary{name={Boltzmann distribution}, description={A probability measure of the microscopic states of a physical system that is composed of classical (i.e. not quantum) particles in thermodynamic equilibrium. This distribution has a density proportional to the factor  $\exp(-E/kT)$, where $E$ is the energy of a state, $k$ is Boltzmann's constant, and $T$ is the absolute temperature.}}

\glossary{name=Detailed balance , description={ An equilibrium where the probability flux of the transitions between any two states is equal in either direction. In population genetics this implies that the numbers of adaptive and deleterious substitutions have to be equal on average.}}

\glossary{name=Entropy , description={A measure of the number of possible configurations of a system. The classical measure of entropy is due to Boltzmann: \(S=-k \log \Omega \), where $\Omega$ is the number (or density) of microscopic states (e.g. allele frequencies) that a system can realize for a given macroscopic state (mean fitness, a quantitative variable, etc.) and $k$ is Boltzmann's constant. Relative entropy is defined as \(S=-\int \psi \log(\psi/\varphi) d\underline{p} \), where the $\underline{p}$ are the microscopic states, and the sum goes over all possible realizations; $\psi$  is the distribution of micro-states, and $\varphi$ is a base or reference distribution (satisfying  \( \varphi = 2N V_{\delta p}\)). However, when $\varphi =$ const. we have Shannon's entropy, which is the form used in statistical physics.}}

\glossary{name=Fisher's information , description={ A measure of how much an infinitesimal change in an unknown parameter $\theta$ affects the likelihood $\psi$ of an observed data set, $p$. Fisher's information is defined as  \(F=\int\psi(p;\theta) \left( \frac{\partial}{\partial \theta}\log[\psi(p;\theta)] \right)^2 dp \). When the parameter $\theta$ is time, Fisher's information describes the amount of information gained through selection. }}

\glossary{name=Fitness flux , description= {A measure of adaptation defined as \(\phi(t)=s(p,t) dp/dt\), where $s$ is the selection coefficient (fitness gradient) and $p$ is the allelic frequency. Geometrically, it is the strength of fitness change (since s is the gradient of fitness, $W$), along the direction of evolution (given by $dp/dt$). The cumulative fitness flux, \( \Phi = \int \phi dt\), is a measure of the total amount of adaptation through the history of a population.}}

\glossary{name=Free fitness , description={The expected gain in log-mean fitness after selection; after an analogy with the free energy of a physical system, that is the amount of work that can be done in a thermodynamic system. Free fitness ($I$) emerges naturally when computing the gain in entropy $S$ after an allele or a trait underwent selection \cite{Iwasa:1988p12}, and has an equivalent expression to free energy, i.e. \(I=\langle \log(\bar{W})\rangle -S/2N\)  (in physics $\langle \log(\bar{W})\rangle$   should be replaced by $\langle E \rangle$ , and $2N$ by $1/kT$; see entry for Boltzmann distribution). } }

\glossary{name= Hill-Robertson interference, description={Interference in the selective sweep of an allele, due to the selective effects at another linked loci. Hill-Robertson interference implies that in the presence of recombination, genotypes with multiple mutations arise easier by recombining existing single mutations than by multiple mutation events. } }

\glossary{name=Path ensemble , description={A formalism of non-equilibrium statistical mechanics and quantum mechanics where the description of the system emphasizes not the states of a population of entities, but rather the distribution of possible stochastic paths that such a population can follow. }}

\glossary{name= Quasispecie, description={Population of replicators (typically asexual) with a high genotypic variability maintained by elevated mutation rates. }}

\glossary{name=Replicator dynamics , description={ Dynamical equations that describe the change in time the frequency $p$ of the different types (in particular genotypes). It has the general form \(dp/dt = p\Delta W + T\) , where $\Delta W$ is the difference between the fitness of the type and the mean fitness, and $T$ are the ``transmission'' terms, that may involve mutation, migration, recombination, etc.} }

\glossary{name=Selective death , description={Failure to survive or reproduce due to differences in genotype.  } }

\glossary{name=Stationary distribution , description= {A probability distribution that does not change in time. This is found from the diffusion equation by setting \(\partial \psi / \partial t = 0\) , and solving the resulting differential equation that is independent of time. A stationary solution might not exist (e.g. if selection is changing in time in particular ways), and if it exists, it might require detailed balance. }}

\glossary{name=Statistical mechanics , description={A mathematical framework explaining the relationship between the macroscopic properties of a system, in terms of the dynamics of the microscopic variables. At equilibrium, it leads to the classical concepts of entropy, free energy, and temperature, for example. Out of equilibrium, these quantities cannot be defined formally, and current research focuses in finding probabilistic measures that apply in general, but are still based on the microscopic dynamics. Based principally on the properties of stochastic processes (e.g. the diffusion equations, or path ensembles), these measures can be applied to the distribution of allele frequencies (e.g. Fisher's information and fitness flux). }}

\glossary{name= Traveling waves, description={Solutions to non-linear differential equations characterized by functions that are of stable shape, and move at a certain velocity either in physical space, or in genetic space. (Traveling waves are also known as \emph{solitons} in the physics and mathematics literature.) }}

\glossary{name=Truncation selection , description= {Scheme where individuals that have traits outside a prescribed range are eliminated. This type of selection is popular in artificial selection.}}


\begin{theglossary}\glogroupB

\gloitem {\glosslabel{glo:Boltzmann distribution}{Boltzmann distribution}}A probability measure of the microscopic states of a physical system that is composed of classical (i.e. not quantum) particles in thermodynamic equilibrium. This distribution has a density proportional to the factor $\exp (-E/kT)$, where $E$ is the energy of a state, $k$ is Boltzmann's constant, and $T$ is the absolute temperature.\relax
\glodelim 
		\glsnumformat{9}\delimT \gloskip \glogroupD

\gloitem {\glosslabel{glo:Detailed balance}{Detailed balance}}An equilibrium where the probability flux of the transitions between any two states is equal in either direction. In population genetics this implies that the numbers of adaptive and deleterious substitutions have to be equal on average.\relax
\glodelim 
		\glsnumformat{9}\delimT \gloskip \glogroupE

\gloitem {\glosslabel{glo:Entropy}{Entropy}}A measure of the number of possible configurations of a system. The classical measure of entropy is due to Boltzmann: \(S=-k \log \Omega \), where $\Omega $ is the number (or density) of microscopic states (e.g. allele frequencies) that a system can realize for a given macroscopic state (mean fitness, a quantitative variable, etc.) and $k$ is Boltzmann's constant. Relative entropy is defined as \(S=-\int \psi \log (\psi /\varphi ) d\underline {p} \), where the $\underline {p}$ are the microscopic states, and the sum goes over all possible realizations; $\psi $ is the distribution of micro-states, and $\varphi $ is a base or reference distribution (satisfying \( \varphi = 2N V_{\delta p}\)). However, when $\varphi =$ const. we have Shannon's entropy, which is the form used in statistical physics. Entropy is also equivalent to the log-likelihood of $\varphi$ (the proposed distribution), and $\psi$ is the sampling probability of the actual distribution.\relax
\glodelim 
		\glsnumformat{9}\delimT \gloskip \glogroupF

\gloitem {\glosslabel{glo:Fisher's information}{Fisher's information}}A measure of how much an infinitesimal change in an unknown parameter $\theta $ affects the likelihood $\psi $ of an observed data set, $p$. Fisher's information is defined as \(F=\int \psi (p;\theta ) \left ( \frac {\partial }{\partial \theta }\log [\psi (p;\theta )] \right )^2 dp \). When the parameter $\theta $ is time, Fisher's information describes the amount of information gained through selection.\relax
\glodelim 
		\glsnumformat{9}\delimT 

\gloitem {\glosslabel{glo:Fitness flux}{Fitness flux}}A measure of adaptation defined as \(\phi (t)=s(p,t) dp/dt\), where $s$ is the selection coefficient (fitness gradient) and $p$ is the allelic frequency. Geometrically, it is the strength of fitness change (since s is the gradient of fitness, $W$), along the direction of evolution (given by $dp/dt$). The cumulative fitness flux, \( \Phi = \int \phi dt\), is a measure of the total amount of adaptation through the history of a population.\relax
\glodelim 
		\glsnumformat{9}\delimT 

\gloitem {\glosslabel{glo:Free fitness}{Free fitness}}The expected gain in log-mean fitness after selection; after an analogy with the free energy of a physical system, that is the amount of work that can be done in a thermodynamic system. Free fitness ($I$) emerges naturally when computing the gain in entropy $S$ after an allele or a trait underwent selection \cite {Iwasa:1988p12}, and has an equivalent expression to free energy, i.e. \(I=\langle \log (\bar {W})\rangle -S/2N\) (in physics $\langle \log (\bar {W})\rangle $ should be replaced by $\langle E \rangle $ , and $2N$ by $1/kT$; see entry for Boltzmann distribution). \relax
\glodelim 
		\glsnumformat{9}\delimT \gloskip \glogroupH

\gloitem {\glosslabel{glo:Hill-Robertson interference}{Hill-Robertson interference}}Interference in the selective sweep of an allele, due to the selective effects at another linked loci. Hill-Robertson interference implies that in the presence of recombination, genotypes with multiple mutations arise easier by recombining existing single mutations than by multiple mutation events. \relax
\glodelim 
		\glsnumformat{9}\delimT \gloskip \glogroupP

\gloitem {\glosslabel{glo:Path ensemble}{Path ensemble}}A formalism of non-equilibrium statistical mechanics and quantum mechanics where the description of the system emphasizes not the states of a population of entities, but rather the distribution of possible stochastic paths that such a population can follow.\relax
\glodelim 
		\glsnumformat{9}\delimT \gloskip \glogroupQ

\gloitem {\glosslabel{glo:Quasispecie}{Quasispecie}}Population of replicators (typically asexual) with a high genotypic variability maintained by elevated mutation rates.\relax
\glodelim 
		\glsnumformat{9}\delimT \gloskip \glogroupR

\gloitem {\glosslabel{glo:Replicator dynamics}{Replicator dynamics}} Dynamical equations that describe the change in time the frequency $p$ of the different types (in particular genotypes). It has the general form \(dp/dt = p\Delta W + T\) , where $\Delta W$ is the difference between the fitness of the type and the mean fitness, and $T$ are the ``transmission'' terms, that may involve mutation, migration, recombination, etc.\relax
\glodelim 
		\glsnumformat{9}\delimT \gloskip \glogroupS

\gloitem {\glosslabel{glo:Selective death}{Selective death}}Failure to survive or reproduce due to differences in genotype. \relax
\glodelim 
		\glsnumformat{9}\delimT 

\gloitem {\glosslabel{glo:Stationary distribution}{Stationary distribution}}A probability distribution that does not change in time. This is found from the diffusion equation by setting \(\partial \psi / \partial t = 0\) , and solving the resulting differential equation that is independent of time. A stationary solution might not exist (e.g. if selection is changing in time in particular ways), and if it exists, it might require detailed balance. \relax
\glodelim 
		\glsnumformat{9}\delimT 

\gloitem {\glosslabel{glo:Statistical mechanics}{Statistical mechanics}}A mathematical framework explaining the relationship between the macroscopic properties of a system, in terms of the dynamics of the microscopic variables. At equilibrium, it leads to the classical concepts of entropy, free energy, and temperature, for example. Out of equilibrium, these quantities cannot be defined formally, and current research focuses in finding probabilistic measures that apply in general, but are still based on the microscopic dynamics. Based principally on the properties of stochastic processes (e.g. the diffusion equations, or path ensembles), these measures can be applied to the distribution of allele frequencies (e.g. Fisher's information and fitness flux).\relax
\glodelim 
		\glsnumformat{9}\delimT \gloskip \glogroupT

\gloitem {\glosslabel{glo:Traveling waves}{Traveling waves}}Solutions to non-linear differential equations characterized by functions that are of stable shape, and move at a certain velocity either in physical space, or in genetic space. (Traveling waves are also known as \emph {solitons} in the physics and mathematics literature.)\relax
\glodelim 
		\glsnumformat{9}\delimT 

\gloitem {\glosslabel{glo:Truncation selection}{Truncation selection}}Scheme where individuals that have traits outside a prescribed range are eliminated. This type of selection is popular in artificial selection.\relax
\glodelim 
		\glsnumformat{9}\delimT 
\end{theglossary}

\begin{figure}[f]
\centering
\includegraphics[scale=1.0]{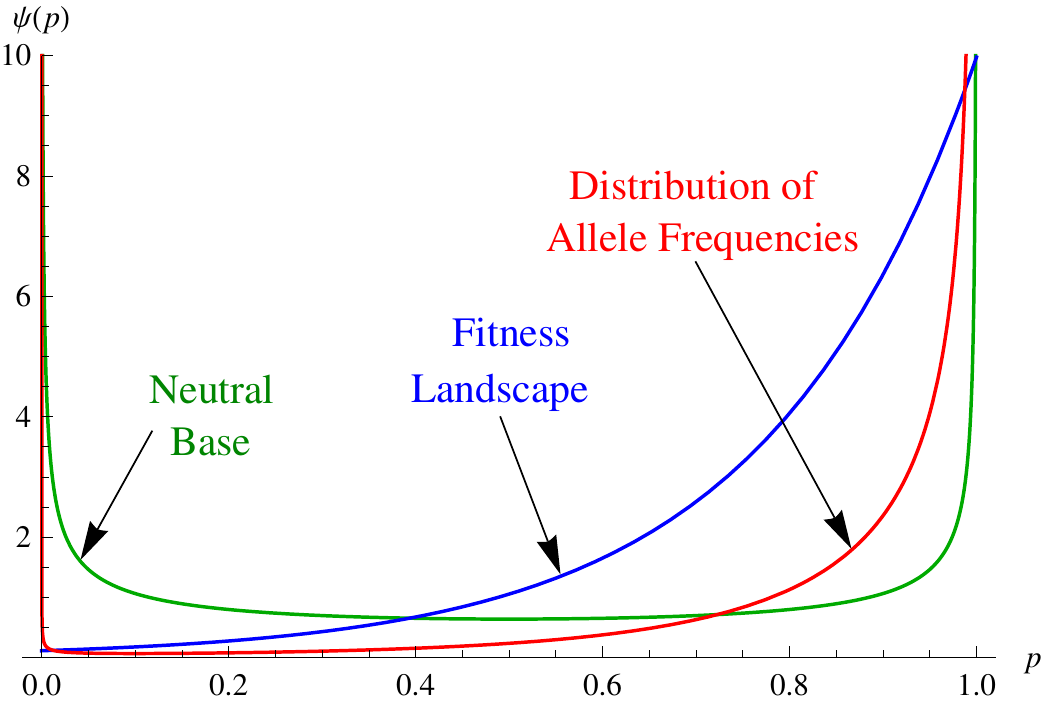}
\caption{In the solution to the diffusion equation, the effects of fitness (blue) combine with neutral factors (green) to give the distribution of allele frequencies (red). The Metropolis-Hastings algorithm has an analogous structure: the acceptance weights (blue) and the random fluctuations (green) combine to give the distribution that is being estimated (red).}
\label{Fig:DESols}
\end{figure}

\begin{figure}[f]
\centering
\includegraphics[scale=0.5]{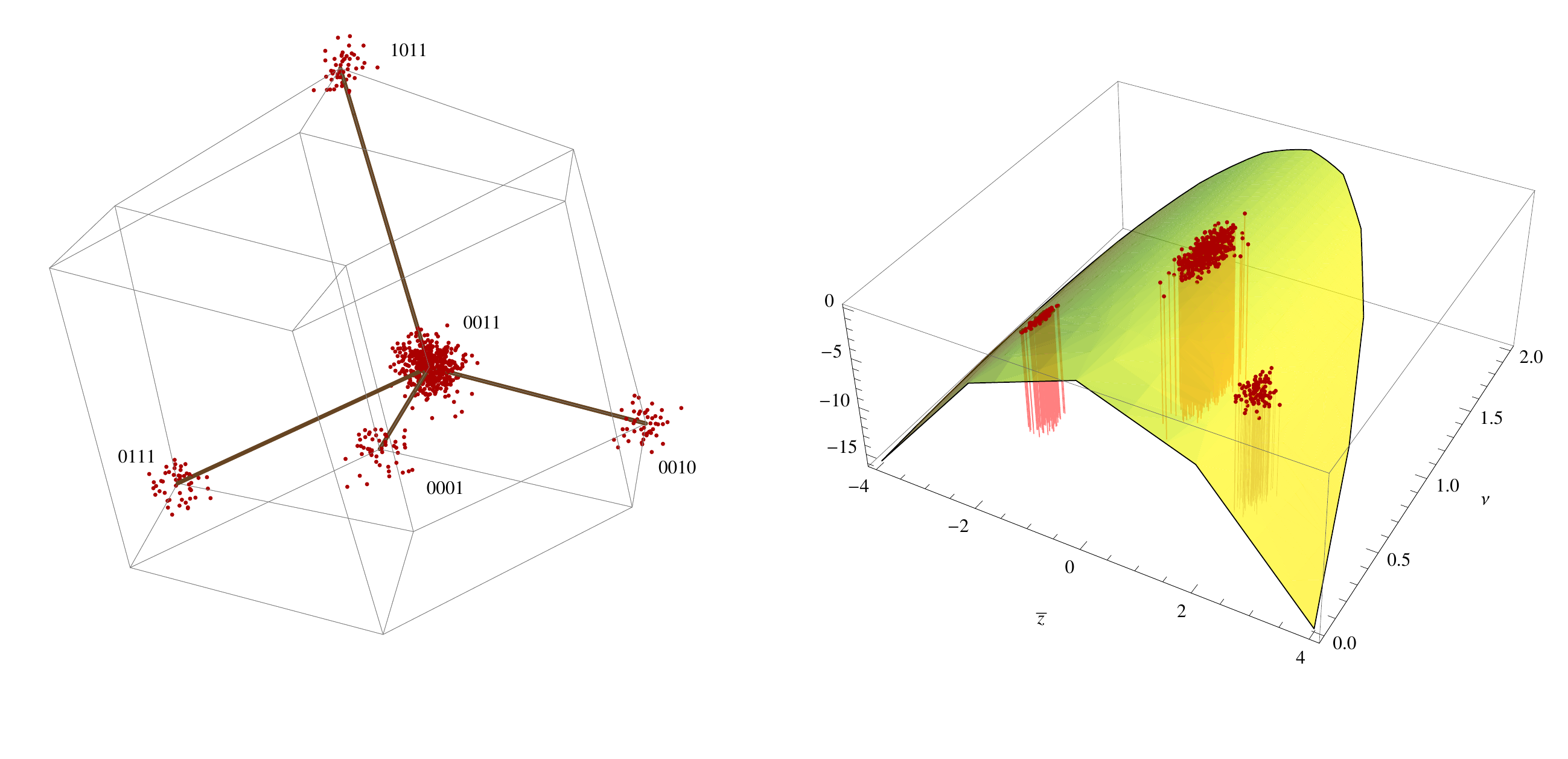}
\caption{Mapping the genetic fitness landscape to a quantitative-trait fitness landscape. Left: different combinations of allele frequencies, lie in a hyperspace (shown only for a projection of 4 loci), where the axes represent the frequency of each allele. In this plot each point represents a population. The dense cloud of points towards the centre is an optimal peak, set at 0011. The other clouds are at sub-optimal adaptive peaks one mutation away from the optimum. However, each genotype determines a trait, and the population is mapped to a space of trait means, $z$, and genetic variance, $\nu$. Thus, mean fitness, trait mean, and genetic variance, although related by the allele frequencies, generate a fitness landscape in quantitative variables (yellow surface, the height indicating log-mean fitness).  The number of variables (degrees of freedom) is collapsed from a hyperspace of an arbitrary number of allele frequencies at each locus to two quantitative variables: trait mean and genetic variance.}
\label{Fig:SimpsonLandscape}
\end{figure}

\begin{figure}[f]
\centering
\includegraphics[scale=1.0]{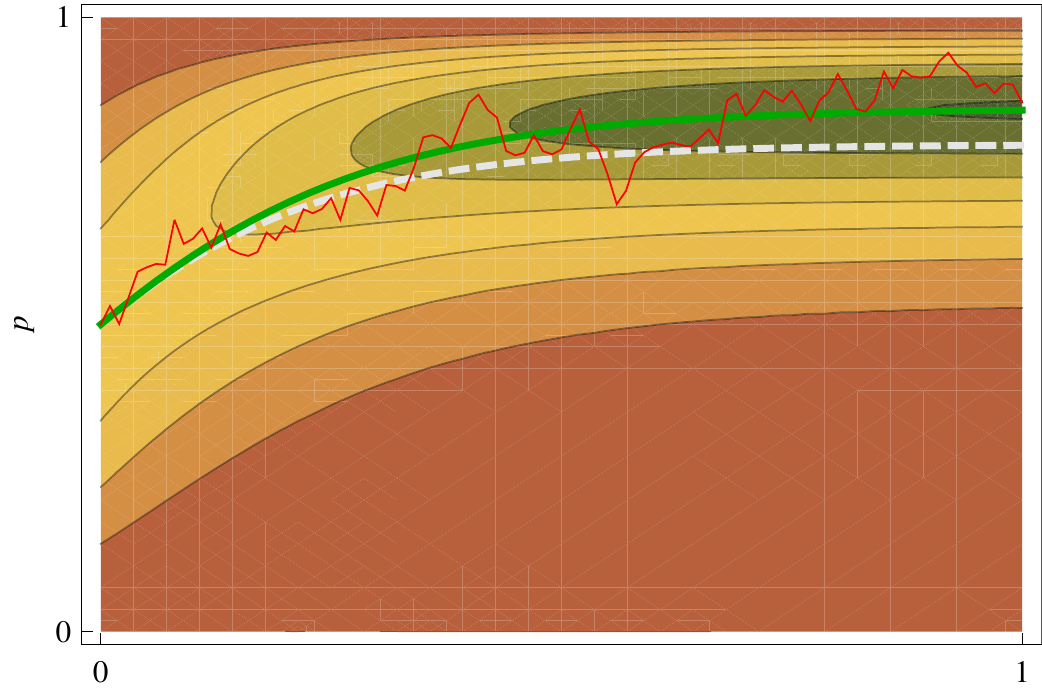}
\includegraphics[scale=1.0]{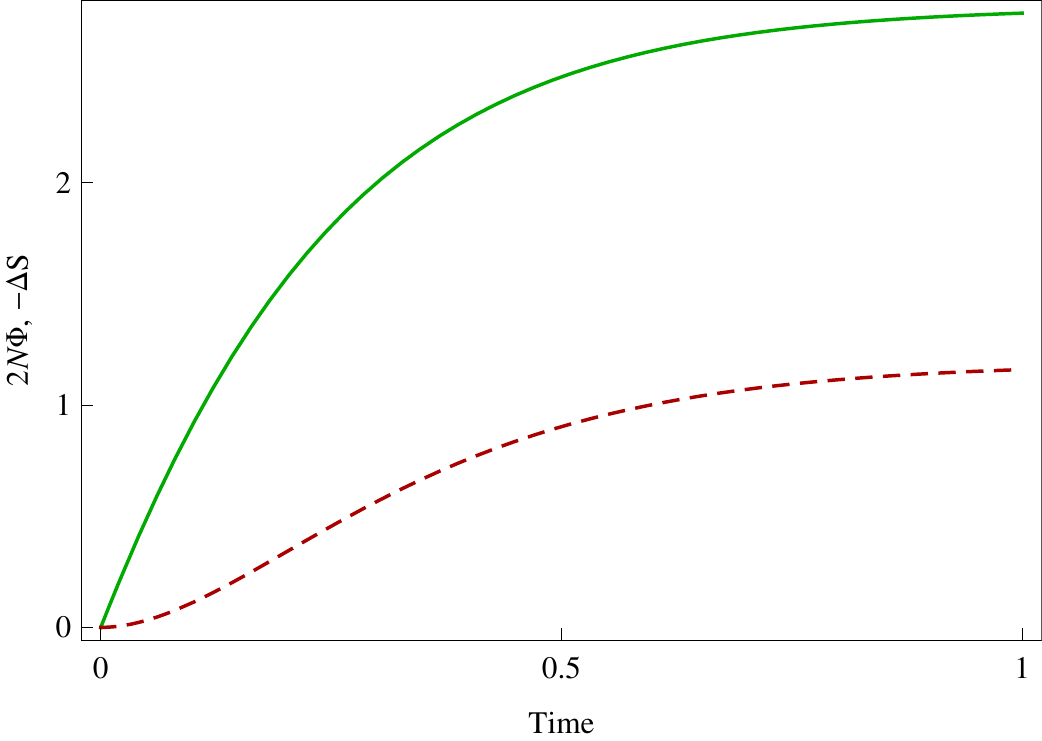}
\caption{The top panel shows the distribution of allele frequencies through time (shown as contour levels). Initially, populations follow the neutral distribution (left axis; $N\mu=0.7$). Directional selection $Ns=2.5$ is then applied, and populations settle to a new distribution (right axis). Any actual realization (red curve) fluctuates stochastically around an optimal one (illustrated with the green curve). The white dashed line is the deterministic solution, shown as a reference. The lower panel shows the fitness flux (upper curve, green) and the decrease in entropy (lower curve, red). When selection changes abruptly, as here, fitness flux is substantially greater than the decrease in entropy. However, if selection were to change slowly, the two would be equal throughout.}
\label{Fig:FitnessFlux}
\end{figure}

\newpage

 \appendix

\section{Evolution and the material basis of heredity}
\label{Box:History}
Early in the last century, Fisher embarked on the mathematical formalization of the Mendelian principles of heredity, following the earlier development of biometrics by Galton, Pearson and Weldon, all with the aim of quantifying evolution by natural selection. Fisher used the diffusion approximation (see \ref{Box:DiffEq}) to describe the evolution of allele frequencies  \cite{Fisher:1922lh}. In 1929, he introduced the Fundamental Theorem of Natural Selection \cite{Fisher30}; comparing it to the second law of thermodynamics, the increase of entropy, he intended the theorem to be an exact result, a ``biological law'' \cite{Price:1972tw,Ewens:1989qo}. Although his comparison with the second law is flawed, it shows how the quantitative approach to heredity was influenced by statistical thermodynamics (indeed, Fisher had studied with E.T. Jeans, a physicist).
That mechanistic basis of evolution, population genetics, was formulated without knowledge of the physical nature of the Mendelian genes, which was still unknown in the 1930's: the structure of DNA was not established until 1953. In the following decades, Delbr{\"u}ck, formerly an astrophysicist, started a collaboration employing ionizing radiation on \emph{Drosophila} to understand the physical nature of the genes, as a working system to try to identify fundamental physical laws that would account for living and non-living matter \cite{Timofeeff-Ressovsky:1935il}. Later, the ingenious Luria-Delbr\"uck experiment proved the basis of Darwinian evolution. They performed a statistical comparison between the number of bacteria developing resistance to lysogenic viruses and its expected distribution, which was derived from a mathematical analysis, an unusual quantitative approach for the biologists of the time \cite{Luria:1943p4720}. Soon after, the quantum theorist Schr\"odinger published \emph{What is life?} \cite{Schroedinger1944}, posing fundamental biological questions in physicists' language -- partly based on Delbr\"uck's discoveries. This book gave strong motivation to the first molecular biologists (among them Perutz, Wilkins, Crick and Watson) to find how DNA transmitted the heritable information to future generations \cite{Judson1998}.
Molecular biology was influenced in large part by the use of physical techniques such as X-ray crystallography to determine biological structures. Evolution, however, while resting on that material basis of DNA, is not explained by it. Indeed, the population genetic framework that we use today was developed prior to the discovery of the structure of DNA, and was not changed by the establishment of molecular biology, since it rests only on MendelÕs laws. However, the theoretical methods that are common to statistical physics and to evolutionary biology give a deeper understanding of the evolutionary consequences of heredity. 

\section{The Diffusion Equation}
\label{Box:DiffEq}
The diffusion equation originated in Bachelier's models of fluctuations in share prices in 1900, and was rediscovered 73 years later as the Black-Scholes formula, disastrously popular amongst economists. Diffusion theory in economics is equivalent to the theory of Brownian motion, devised by Einstein to explain random molecular collisions, and soon after extended in physical applications \cite{Davis:2006mb}. Fisher \cite{Fisher:1922lh} compared Mendelian genetics to ``the theory of gases'' and introduced the diffusion methods for the allele frequencies. Kolmogorov \cite{Kolmogorov:1931ye} gave a more formal approach to selection and drift. Kimura \cite{Kimura:1964p5898} later extended this formalism to non-equilibrium cases.
	For population genetics, the diffusion equation is a rather convenient representation of evolution of finite populations where genetic drift is present. We could choose to model the change in allele frequencies directly, in what is known a Wright-Fisher process. But genetic drift evolves stochastically, making the outcomes of evolution unpredictable. The diffusion equation gauges these outcomes in a probabilistic way, describing the distribution of allele frequencies at each time. (A third way to describe an evolving populations is to use the whole history as a random variable instead of the allele frequencies at a time, in what is known as a path ensemble; see \ref{Box:PathEnsemble}). In short, the diffusion equation is a partial differential equation describing the change in time of the probability density $\psi$  of the allele frequencies $p$, namely
  \[\frac{\partial \psi}{\partial t}=-\frac{\partial}{\partial p}(M_{\delta p} \psi)+\frac{1}{2}\frac{\partial^2}{\partial p^2}(V_{\delta p}\psi) ~,\]
where \(M_{\delta p} \)  are the deterministic factors, due to selection, mutation, migration, etc. and \(V_{\delta p}\)  is the variance of the fluctuations by drift usually of the form  \(p(1-p)/2N\). Making the left-hand side equal to zero, leads to the stationary solution derived by Wright by other means [16]:
  \[\psi =C\bar{W}^{2N} \left[p(1-p)\right]^{4N\mu-1} .\]
For details of the derivation see \cite{Kimura:1964p5898}. In particular, the term \(\bar{W}\) defines the ``fitness landscape'' which can be thought as a surface in the space of allele frequency (Fig. \ref{Fig:DESols}), or in the quantitative variables (Fig. \ref{Fig:SimpsonLandscape}).

The diffusion equation, the coalescent process, and the path ensemble all describe the same process and are mathematically equivalent. Each has different advantages and limitations; whereas a stochastic differential equation, the diffusion equation and the path ensemble do not require detailed balance, the stationary distribution above, does. Yet, this solution is exact, quite general and relatively simple.

\section{Path ensembles and fitness flux}
\label{Box:PathEnsemble}
A path ensemble considers all the possible histories of a population between two fixed states $p_0$ and $p_T$ at times $0$ and $T$. In this description, each history is the variable being described. The probability of a particular trajectory $\rho(t)$ is proportional to the factor \(\exp\left[ -N\int \left(\frac{dp}{dt}- M_{\delta p} \right)^2\frac{dt}{p(1-p)}\right]~,\) 
where the allele frequencies $p$ are evaluated at each  point of the history $\rho(t)$. Here, $M_{\delta p}$ is the same factor in the diffusion equation --suggesting the connection between the two methods. The path integral can be understood as a sum if the history is sampled at discrete times, \(\rho=\{\rho_0, \rho_1, \ldots, \rho_T\}\). Notice that because the integral is always positive, if it achieves a minimum for a given history, then that history has the highest probability. To understand the meaning of the integral we may develop the binomial expression inside the integral into three terms: \(F-2\phi+\nu\), and consider the case of selection,  $M_{\delta p}=p(1-p)s$ ; $F$ is  Fisher's information, \(\phi=\frac{M_{\delta p}}{p(1-p)}\frac{dp}{dt}=s\frac{dp}{dt}\)  is the fitness flux, and $\nu$ is the additive genetic variance in fitness. Thus the histories occur as a compromise between minimizing Fisher's information and genetic variance --both regarded as measures of the speed of adaptation-- and maximizing the fitness flux. 

Fitness flux is a measure of adaptation of beneficial alleles \cite{Mustonen:2010p5306} the cumulative flux \(\left(\Phi=\int \phi dt \right)\) of a population history is the equivalent measure to the fitness of a population (if we think of successive substitutions, it is the total of all the selection coefficients associated with each substitution). The expectation of cumulative fitness flux  is necessarily greater than the reduction in entropy between the initial and final equilibrium states (which can be understood as the information gained by the population) [49]: \(2N\langle \Phi \rangle\geq -\Delta S\). That is, it takes a certain amount of selection (measured precisely by the fitness flux) to move the allele frequency distribution away from its neutral state (as measured by the decrease in entropy). This result is quite generally valid, and is not restricted to, say, constant selection. Moreover, if selection changes slowly so that the distribution stays close to the stationary state, then \(2N\langle \Phi \rangle = -\Delta S\) ; such changes are termed ``reversible''. 
For example, assume that the allele frequencies initially follow a neutral distribution (mutation-drift balance). Suddenly, directional selection is applied so that \(\bar{W}=\exp(sp)\), and loci move toward a new distribution under selection and drift. The fitness flux is then substantially greater than the decrease in entropy (Fig. \ref{Fig:FitnessFlux}). If, on the other hand, selection were increased very slowly, eventually to reach the same strength, the net fitness flux would necessarily be much smaller, and equal to the decrease in entropy (lower curve in Fig. \ref{Fig:FitnessFlux}). The fitness flux method is surprisingly general. However, its relation with quantities that might actually constrain the extent of selection. In particular, the additive variance in fitness is proportional to \(s^2p(1-p)\) ; we see that the additive genetic variance in fitness is just twice the fitness flux, when that includes only the change in allele frequency due to selection,  \(\Delta_s p=s p(1-p)\). Further understanding could emerge relating the decrease in entropy due to selection to the additive genetic variance in fitness \cite{Barton:2000p5970}.

Last, it is relevant that fitness flux presents an extension of Fisher's Fundamental Theorem of Natural selection \cite{Fisher30}: it considers not only the change due to selection, but also the effects of drift, and unlike Fisher's theorem, the fitness flux theorem holds also for weak selection ($Ns\sim 1$) \cite{Mustonen:2010p5306}.

\newpage






\end{document}